\documentclass[draftcls, 11pt, onecolumn]{IEEEtran}
\usepackage{amssymb, amsmath,amsthm, graphicx, paralist,subfigure,cite}
\usepackage{algorithm2e}
\usepackage{color}
\newtheorem{lem}{Lemma}
\newtheorem{theorem}{Theorem}
\newtheorem{defn}{Definition}

\newtheorem{cor}{Corollary}

\def\mc{\mathcal}
\DeclareMathOperator*{\argmin}{argmin}

\begin{document}

\title{Minimal Driver Nodes for Structural Controllability of Large-Scale Dynamical Systems:  Node Classification}

\author{Mohammadreza Doostmohammadian$^\ast$
	
\thanks{
	$^\ast$ Mechanical Engineering Department, Semnan University, Semnan, Iran, \texttt{doost@semnan.ac.ir}}}
\maketitle

\begin{abstract}
This paper considers the problem of minimal control inputs to affect the system states such that the resulting system is structurally controllable. This problem and the dual problem of minimal observability are claimed to have no polynomial-order exact solution and, therefore, are NP-hard. %\footnote{NP-hard problems are claimed to have no solution in time-complexity upper-bounded by a polynomial function of input parameters.}. 
Here, adopting a graph-theoretic approach, this problem is solved for general nonlinear (and also structure-invariant) systems and a P-order solution  is proposed. In this direction, the  dynamical system is modeled as a directed graph, called \textit{system digraph}, and two types of graph components are introduced which are tightly related with structural controllability. Two types of nodes which are required to be affected (or driven) by an input, called \textit{driver nodes}, are defined, and minimal number of these driver nodes are obtained.  Polynomial-order complexity of the given algorithms to solve the problem ensures applicability of the solution for analysis of large-scale dynamical systems. {The structural results in this paper are significant as compared to the existing literature which offer approximate and computationally less-efficient, e.g. Gramian-based, solutions for the problem, while this paper provides exact solution with lower computational complexity and applicable for controllability analysis of nonlinear systems. }

\end{abstract}

\section{Introduction}\label{sec_introduction}

\IEEEPARstart{C}{ontrollability} describes the ability of inputs to drive the dynamical system from any initial state to the desired final state in finite time, while the dual concept of  observability is a measure of how well internal states of the dynamical system can be inferred by external state measurements in finite time. Controllability and observability have been topics of interest in analysis of variety of systems, including smart grid and power systems \cite{camsap11,heussen2011unified,haynes2002domination}, biological systems \cite{liu_pnas,orosz2010controlling}, chemical systems \cite{haber2017state}, and even social systems \cite{jstsp14,pequito_gsip,doostmohammadian2019cyber}.  Due to the large size of these systems, the identification of a small fraction of their states  to drive the entire system  around the state space poses an important problem. This problem, sometimes referred to as the minimal controllability problem \cite{olshevsky2014minimal}, is the main focus of this paper and  finds different applications in leader selection\cite{jafari2011leader,ji2006leader}, cooperative control \cite{zamani2009structural,jadbabailinmorse03}, formation stabilization \cite{cai2010formation,chen2005formation}, synchronization \cite{porfiri2008criteria,tang2014general}, complex network controllability \cite{Liu_nature,sorrentino2007controllability}, {consensus \cite{shang2012finite,shang2017finite,scientia,shang2016combinatorial,shang2014continuous}}, etc. {As an example, applying minimal controllability in consensus (or finite-time consensus) \cite{shang2012finite,shang2017finite,scientia,shang2016combinatorial,shang2014continuous}, one can find the minimal number of \textit{leader agents} to drive the multi-agent system towards reaching consensus value. }

\textit{Related literature:} different aspects of minimal control input problem are considered in the literature. The problem of choosing the minimal set of inputs for optimizing the control energy to drive the dynamical system to the desired state is known to be NP-hard  and therefore $\log n$-approximate solution is proposed \cite{tzoumas2015minimal}. Further, in \cite{summers2014optimal} it is shown that for some specific controllability-Gramian related  metrics the problem, under the special case, yields modular set functions and can be efficiently optimized. The problem of finding the minimal set of inputs to make the system controllable is claimed to be NP-hard \cite{pequito2015complexity,olshevsky2014minimal} and $\log n$-approximate solution is developed in \cite{olshevsky2014minimal}. The dual problem of optimal sensor selection for system observability is claimed to be NP-hard as well and $n^2$-approximate solution is proposed \cite{jiang2003optimal}. Similarly, in \cite{pequito_gsip} it is claimed that the problem of minimal number of \textit{information gatherers} (observer nodes) for observability of a social system is NP-hard in general. In \cite{Liu_nature} the minimal set of driver nodes (the nodes to be injected by input) in a Strongly-Connected (SC) network is determined as the minimal number of unmatched nodes, however, for general (non-SC) networks no result is derived.
The other less-related topic is optimal link/node addition to improve structural controllability \cite{chen2019minimal,commault2013input,jafari2011leader}.  

{An interesting approach is developed in  \cite{alcaraz2013structural,alcaraz2014recovery} to restore/maintain controllability of networks in the presence of adversarial attacks or failure that remove some nodes/links. The authors apply a novel computationally efficient sub-optimal approximation for the restoration of the \textit{Power Dominating Set (PDS)}\footnote{{A set $\mc{S}_G$ is a \textit{Dominating Set} of the graph  $\mc{G}=\{\mc{V},\mc{E}\}$ if every node in $\mc{V}\setminus \mc{S}_G$ has a neighbor node in $\mc{S}_G$. Further define $\mc{S}_G$ as a Power Dominating Set (PDS) if every link/node in $\mc{G}$ can be observed by $\mc{S}_G$ \cite{haynes2002domination}. Finding the minimal PDS is known to be generally NP-complete \cite{alcaraz2014recovery,haynes2002domination}, while P-order approximations are given, e.g., for graphs of bounded treewidth \cite{kneis2006parameterized}. The PDS is first applied in the observability of electrical power networks as the optimization problem of measuring all nodes by placing as few PMUs as possible \cite{kneis2006parameterized,haynes2002domination}.}}. The authors apply their restoration strategy over different random models including \textit{Erdos-Renyi (ER)}, \textit{Small-World (SW)}, and \textit{Scale-Free (SF)} graphs \cite{alcaraz2013structural} and also provide a thorough complexity analysis of their method \cite{alcaraz2014recovery}. Further, this restoration strategy is applied for cloud-based monitoring of Cyber-Physical-Systems (CPS) \cite{alcaraz2018cloud}, where a local attack detection technique based on the opinion dynamics is adopted to find the topological changes in the underlying network and activate a resilience process in case of detecting a threat. \cite{alcaraz2018cloud} further shows that the proposed recovery approach considering structural controllability can reach optimal values in linear times. In general, \cite{alcaraz2013structural,alcaraz2014recovery,alcaraz2018cloud} propose a strategy to reconstruct the PDS in case of topological changes in, e.g., power networks; however, the results can be extended to other types of networks.}

Overall, what missing from the literature is a \textit{polynomial-order} solution of the minimal driver node problem for general dynamical systems with possible non-SC system digraphs. {One specific application of such solution is in the controllability (or dual case of observability) of smart grid and power networks \cite{camsap11,heussen2011unified,haynes2002domination,kneis2006parameterized}. Such networks are of large-scale and not SC in general and, therefore, the existing computationally less-efficient approximate solutions are not practical. Similar reasoning holds for biological system applications \cite{liu_pnas,orosz2010controlling}.}

\textit{Contribution:} motivated by recent advances in structural analysis, a graph-theoretic approach is adopted in this paper to solve the problem of minimal driver nodes for controllability. In this direction, it is shown that the driver nodes  are tied with two components in  the system digraph: (i) the dilation and (ii) the child SCC, and two types of driver nodes are introduced on the system digraph:  Type-I to recover input-connectivity of state nodes, and Type-II to recover rank condition for structural controllability. The minimal set of driver nodes is achieved by finding the shared nodes between these two components in the system digraph.
The significance of the contribution of this paper is stated in the following:
\begin{itemize}
	\item Different realizations of minimal controllability problem are claimed to be NP-hard \cite{tzoumas2015minimal, summers2014optimal, pequito2015complexity,olshevsky2014minimal, jiang2003optimal,pequito_gsip}, and therefore, approximate (and not exact) solutions are suggested  to solve the problem. However, a polynomial-order exact solution of complexity $\mc{O}(n^{2.5})$ is proposed here.
	\item {Only Type-II driver nodes are considered in \cite{Liu_nature} for SC directed networks. However, in this work, the Type-I driver nodes are introduced and the problem of minimal driver nodes is solved for general systems with possible non-SC system digraphs. In general, here it is proved that the minimal number of driver nodes for general systems might be more than what is claimed in \cite{Liu_nature}.}
	\item In contrast to linear controllability considered in \cite{tzoumas2015minimal, summers2014optimal,pequito2015complexity,olshevsky2014minimal,jiang2003optimal,pequito_gsip},  this paper similar to \cite{Liu_nature} generalizes the solution for both linear systems and structure-invariant systems prevalent in linearization of nonlinear systems.
	\item In this paper, the algorithm complexity  using graph-theoretic  approach is $\mc{O}(n^{2.5})$, while the complexity of Gramian-based numerical analysis \cite{tzoumas2015minimal, summers2014optimal} is (at least) $\mc{O}(n^{3})$. Lower computational complexity ensures better efficiency and practicality in large-scale applications.
\end{itemize}

\section{Problem Statement}\label{sec_prob}
In this paper the underlying dynamical system is modeled by coupled first-order {Ordinary Differential Equations} (ODEs) in the form,
\begin{eqnarray} \label{eq_nonlin_sys}
\dot{x} =  f(x)+Bu
\end{eqnarray}
where $x \in\mathbb{R}^n$ is the state vector, $f:\mathbb{R}^n \rightarrow \mathbb{R}^n $ is a nonlinear continuously differentiable function, $u \in \mathbb{R}^N$ is the input, and $B \in \mathbb{R}^{n \times N}$ is the input matrix.
For controllability analysis, one may consider the linearized model of the dynamics~\eqref{eq_nonlin_sys},
\begin{eqnarray} \label{eq_lin_sys}
\dot{x} = Jx+Bu
\end{eqnarray}
where $J \in \mathbb{R}^{n \times n}$ is the Jacobian matrix defined as $J_{ij}= \frac{\partial f_i}{\partial x_j}$ \cite{nonlin}. If $f_i$ is a function of $x_j$ the entry $J_{ij}$ is nonzero, where the exact numerical value is determined by the linearization point. Having the function $f$ to be time-invariant, the structure of the Jacobian matrix $J$, i.e. its zero-nonzero pattern, is fixed for all linearization points.

The nonlinear and linear model of the dynamical system are typically represented as a graph, known as the system digraph. In the system digraph $\mc{G} = \{\mc{V},\mc{E}\}$, node $\mc{V}_i \in \mc{V}$ represents state $x_i$ of the dynamical system. If the entry $J_{ij}$ is nonzero, implying that $f_i$ is a function of $x_j$, there is a link $(\mc{V}_i,\mc{V}_j) \in \mc{E}$ ($\mc{V}_i \rightarrow \mc{V}_j$) in the system digraph $\mc{G}$. Therefore, the system digraph $\mc{G}$ represents the zero-nonzero pattern (or the structure) of the Jacobian matrix $J$. It is known that many properties of the dynamical system are inherent in the structure of its system digraph\footnote{It should be noted that another important example on the zero-nonzero pattern (structure) of the system digraph is the calculation of number of spanning trees, see \cite{shang2016number} for more details.} \cite{woude:03}, including system controllability and observability. Such properties only rely on the zero-nonzero pattern of the Jacobian matrix and are irrespective of the exact numerical values of $J$, and therefore are called \textit{generic}. 
%The numerical values (operating points) for which the generic property does not hold lie on an algebraic subspace with zero Lebesgue measure, and therefore it is typical to say that these properties hold for \textit{almost all} values of system parameters \cite{woude:03}. 
In this direction, the structural observability and controllability are checked based on graph-theoretic methods instead of numerical Gramian-based analysis or PBH test. {It is known that the controllability of a nonlinear system is structurally similar to its linearized model. In fact, having the linearized model to be controllable at every linearization point implies the controllability of the nonlinear model \cite{nonlin}. In other words, since the structural controllability implies the controllability over a continuum of linearized points, the controllability results in this paper are valid for nonlinear case.
}
%Knowing that the controllability of the nonlinear model \eqref{eq_nonlin_sys} can be structurally derived by its linear counterpart model \eqref{eq_lin_sys} (see \cite{nonlin} for more information), 
The main theorem on structural controllability is stated below:

\begin{theorem} \label{thm_scont}
	\cite{lin} A dynamical system in the form \eqref{eq_nonlin_sys}-\eqref{eq_lin_sys} is structurally controllable if and only if in its system digraph $\mc{G}$ the following conditions  hold:
	\begin{enumerate} [(i)]
		\item Every state node $\mc{V}_i$ is the end node of a path initiated from an input or a driver node.
		\item There is a \textit{disjoint} family of cycles and input-connected paths covering all state nodes in $\mc{V}$.
    \end{enumerate}
\end{theorem}

The condition (i) in the above theorem is referred to as the \textit{input-connectivity condition} and condition (ii) is referred to as the \textit{rank condition}.
The main goal in this paper is to determine the minimal number of driver nodes to be affected by an input such that the system digraph $\mc{G}$ is structurally controllable. Without loss of generality assume that each input drives one state node, i.e. each column of the input matrix contains only one nonzero entry. Let $\mc{B}$ denote the zero-nonzero pattern of the input matrix $B$, $\mc{J}$ denote the zero-nonzero pattern of the Jacobian matrix $J$, and $|\mc{B}|_0$ define the number of nonzero entries of $\mc{B}$. The main problem can be formulated as follows:

\begin{equation} \label{eq_prob1}
\begin{aligned}
\displaystyle
\argmin
\limits_{\mc{B}} ~~ & |\mc{B}|_0 \\
\text{s.t.} ~~ & (\mc{J},\mc{B})~\mbox{structural-controllability.}
\end{aligned}
\end{equation}

The goal of this paper is to find a polynomial-order solution for problem~\eqref{eq_prob1} with the assumption that matrix $\mc{J}$ has a fixed structure. This problem for general non-SC systems is claimed to be NP-hard \cite{pequito2015complexity,olshevsky2014minimal, jiang2003optimal,pequito_gsip} and, therefore,  approximate solutions are proposed which are not computationally efficient. In this paper, a P-order exact solution with low computational complexity is proposed applicable in  large-scale systems.

\textit{Assumptions:} In this paper, the network is structurally static and time-invariant while the link weights may change in time. This implies that the system adjacency matrix is fixed-structure while the entries may vary in time. In nonlinear systems this means that the nonlinear characteristic function is time-invariant and, therefore, the structure of Jacobian is fixed.

\section{Related Graph-Theoretic Concepts}\label{sec_graph}
{The concept of system-digraph is used to structurally model dynamical systems \cite{woude:03,jstsp14,shang2016combinatorial,shang2014continuous}.} This section introduces the main graph-theoretic notions related to  structural controllability of system digraphs, namely, dilations and SCCs.  An introduction on SCC classification for structural observability can be found in the previous work by author \cite{asilomar11}.

\subsection{Dilations} \label{sec_dil}
A dilation is defined as a component in the graph in which there are less nodes dilated (or linked) to more other nodes. Let $|.|$ denotes the cardinality of a set and $\mc{N}(.)$ represents the set of neighbors of a node or a subset of nodes, e.g., for the set $\mc{D}_i$ we have $\mc{N}(\mc{D}_i)=\{\mc{V}_j|(\mc{V}_j,\mc{V}_i) \in \mc{E}, \mc{V}_i \in \mc{D}_i \}$.  Then, the \textit{dilation set} is defined as follows:

\begin{defn}\label{def_dilation}
	\cite{Liu_nature} A dilation is a component in which there is a subset of nodes $\mc{D}_i \subset \mc{V}$ such that $|\mc{N}(\mc{D}_i)|<|\mc{D}_i|$, where $\mc{D}_i$ is the dilation set.
\end{defn}

The procedure of finding the dilations and dilation sets are better defined over bipartite graphs. Define the bipartite representation of the system digraph $\mc{G} = (\mc{V},\mc{E})$ as $\Gamma=(\mc{V}^+,\mc{V}^-,\mc{E}_\Gamma)$ with two disjoint set of nodes $\mc{V}^+$ and $\mc{V}^-$ where every link in $\mc{E}_\Gamma$ starts from a node in $\mc{V}^+$ and ends in~$\mc{V}^-$. In $\Gamma$ the node sets are defined as $\mc{V}^+=\mc{V}$ and $\mc{V}^-=\mc{V}$. Further, $\mc{E}_\Gamma$ is the collection of the links defined as~$\{(\mc{V}_j^+,\mc{V}_i^-)|(\mc{V}_j,\mc{V}_i) \in \mc{E}\}$. A \textit{matching}, $\underline{\mc{M}}$, in $\Gamma$ is the set of mutually disjoint links that share no start node in~$\mc{V}^+$ and no end node in~$\mc{V}^-$. A matching with maximum size is called the \textit{maximum matching}, $\mc{M}$. Denote  by~$\partial \mc{M}^-$ the \textit{matched} nodes as the nodes in~$\mc{V}^-$ incident to~$\mc{M}$ and define the \textit{unmatched} nodes as~$\delta\mc{M} = \mc{V}^- \backslash \partial \mc{M}^-$. In fact, a node is matched if it is an ending node of a link in $\mc{M}$; otherwise the node is \textit{unmatched}.
A maximum matching $\mc{M}$ can be obtained from a matching $\underline{\mc{M}}$ in the \textit{auxiliary graph}~$\Gamma^{\underline{\mc{M}}}=(\mc{V}^+,\mc{V}^-,\mc{E}^{\mc{M}}_{\Gamma})$. Having a matching $\underline{\mc{M}}$, the graph $\Gamma^{\underline{\mc{M}}}$ has the same set of nodes as in  $\Gamma$, while the set of links $\mc{E}^{\mc{M}}_{\Gamma}$ is obtained from $\mc{E}_{\Gamma}$ by reversing the direction of the links in~$\underline{\mc{M}}$  and keeping the direction of other links in~$\mc{E}_{\Gamma} \backslash \underline{\mc{M}}$. Denote by~$\mc{Q}_{\underline{\mc{M}}}$ a \textit{${\underline{\mc{M}}}$-alternating path}  as sequence of links alternating between matched links~$\mc{M}$ and unmatched links in~$\mc{E}_{\Gamma} \backslash \underline{\mc{M}}$. Such sequence starts from a node in~$\delta \underline{\mc{M}}$ with a  link in~$\mc{E}_{\Gamma} \backslash \underline{\mc{M}}$  and every second link in~$\underline{\mc{M}}$. Denote by ~$\mc{P}_{\underline{\mc{M}}}$ an
\textit{${\underline{\mc{M}}}$-augmenting path} as an alternating path starting and ending in~$\delta \underline{\mc{M}}$. In fact, the ${\underline{\mc{M}}}$-augmenting path is used to find the maximum matching from a simple matching (see Algorithm~\ref{alg_dilation}). Having a maximum matching $\mc{M}$, choose a node~$\mc{V}_i \in \delta \mc{M}$ in $\Gamma^\mc{M}$ and find the set of all nodes in~$\mc{V}^-$ reachable by  alternating paths~$\mc{Q}_{\mc{M}}$ from~$\mc{V}_i$. This resultant set is a dilation set. Denote by $\mc{D}$  the set of all dilation sets, i.e. $\mc{D}=\{\mc{D}_1,...,\mc{D}_l\}$. This procedure of finding a maximum matching $\mc{M}$ and consequently finding the dilation sets $\mc{D}_i$ is summarized in Algorithm~\ref{alg_dilation}.
\begin{algorithm} \label{alg_dilation}
	\textbf{Given:} System digraph $\mc{G}$
	
	\KwResult{Dilations $\{\mc{D}_1,...,\mc{D}_l\}$}
	Construct $\Gamma=(\mc{V}^+,\mc{V}^-,\mc{E}_\Gamma)$\;
	Find a matching $\underline{\mc{M}}$ \;
	Construct $\Gamma^{\underline{\mc{M}}}$ by reversing the links of $\underline{\mc{M}}$ in $\Gamma$\;
	Find~$\partial {\underline{\mc{M}}}^-$ as the nodes in~$\mc{V}^-$  incident to~${\underline{\mc{M}}}$ \;
	Find~$\delta{\underline{\mc{M}}} = \mc{V}^- \backslash \partial {\underline{\mc{M}}}^-$\;
	Find~$\mc{Q}_{\underline{\mc{M}}}$ as a sequence of links starting from an unmatched node in~$\delta {\underline{\mc{M}}}$ and every second link in~${\underline{\mc{M}}}$\;
	Find~$\mc{P}_{\underline{\mc{M}}}$ as a $\mc{Q}_{\underline{\mc{M}}}$ with start and end node in $\delta {\underline{\mc{M}}}$\;
	\While{ $\mc{P}_{\underline{\mc{M}}}$ exist}{
		\For{nodes in $\delta \underline{\mc{M}}$}{
			Find $\mc{P}_{\underline{\mc{M}}}$ \;
			$\underline{\mc{M}} = \underline{\mc{M}} \oplus \mc{P}_{\underline{\mc{M}}}$ \;
		}
	}
	
	Construct $\Gamma^{\mc{M}}$ for the maximum matching ${\mc{M}}$\;
	Find ~$\partial {\mc{M}}^-$ and $\delta {\mc{M}} = \mc{V}^- \backslash \partial \mc{M}^-$\;
	Find~$\mc{Q}_{\mc{M}}$ for the maximum matching~${\mc{M}}$\;	
	\For{nodes in $\delta \mc{M}$}{
		Find $\mc{Q}_{\mc{M}}$ in $\Gamma^{\mc{M}}$ \;
		Put all nodes in $\mc{V}^-$ reachable by $\mc{Q}_{\mc{M}}$ in $\mc{D}_i$\;}
	
	\textbf{Return} $\mc{D}_i, i=\{1,...,l\}$\;\
	
	\caption{The first loop in the algorithm renders the maximum matching~$\mc{M}$, and the second loop in the algorithm renders the dilation set for each unmatched node in~$\mc{M}$. Note that $\oplus$ represents the XOR operator in set theory. {XOR (Exclusive OR) is an operation that outputs true (or $1$) only when logical inputs differ.}}
\end{algorithm}
%We illustrate the above procedure of finding dilations by examples in Section~\ref{sec_example}.

\subsection{Child SCCs} \label{sec_scc}
Define a Strongly Connected Component (SCC), denoted by $\mc{S}_i$, as the \textit{largest} set of nodes in the system digraph $\mc{G}=\{\mc{V},\mc{E}\}$ in which every two nodes are connected via a directed path, i.e. $\forall \{\mc{V}_i,\mc{V}_j \} \in \mc{S}$ we have $\mc{V}_i \xrightarrow{path} \mc{V}_j$ and $\mc{V}_j \xrightarrow{path} \mc{V}_i$. Further, classify the SCCs in terms of their input-connectivity as \textit{child} and \textit{parent} SCCs; a child SCC, denoted by $\mc{S}^c_i$, is a SCC with no incoming link from nodes not belonging to itself, i.e. $\forall \mc{V}_i \in \mc{S}^c_i, \nexists  (\mc{V}_j,\mc{V}_i)\in \mc{E},\mc{V}_j \notin \mc{S}^c_i$. A SCC that is not a child is called a parent SCC. Further, define the partial order of SCCs by $\prec$. $\mc{S}_i \prec \mc{S}_j$ implies that there is (at least) one link (or a directed path) from the nodes in $\mc{S}_i$ to nodes in $\mc{S}_j$. This implies that for every child SCC $\mc{S}^c_i$ there is no other SCC $\mc{S}_j$ such that $\mc{S}_j \prec \mc{S}^c_i$. Denote by $\mc{S}^c$ the set of all child SCCs, i.e. $\mc{S}^c=\{\mc{S}^c_1,...,\mc{S}^c_p\}$. The classification of SCCs and their partial order can be defined by applying the well-known \textit{DFS} algorithm \cite{algorithm}.

The DFS algorithm starts from a \textit{root node} in the digraph and explores all nodes in the digraph. This so-called root node can be chosen as any arbitrary node in the digraph, where the algorithm starts the search from it. From this root node the algorithm follows the links in the digraph to reach nodes not visited before. As the algorithm goes deeper in the graph, a new root node may be chosen if not all nodes are visited from the previous root node. The algorithm ends as all nodes in the digraph are visited. The algorithm keeps track of the predecessor of each visited node. The graph associated with this predecessor is a \textit{tree} graph called \textit{DFS forest}\footnote{{A (DFS) tree contains all nodes reachable from the root node. A complete DFS exploring the entire graph (and not only the part reachable from a given root node) builds up a collection of trees, or a forest, called a DFS forest. A recent work  \cite{shang2016likelihood} develops the idea of graph likelihood in \textit{forest graphs} to grow sparse graphs to model real-world networks. }}.  
For each node $\mc{V}_i$ the following attributes are saved: (i) $\mc{V}_i.\pi$ denoting the predecessor of node $\mc{V}_i$, (ii) $\mc{V}_i.u$ denoting a boolean variable which is false if the algorithm has not visited $\mc{V}_i$ yet and true otherwise, and (iii) $\mc{V}_i.s$ and $\mc{V}_i.e$ denoting the start time and the end time a node is visited. The modified DFS algorithm is summarized in the Algorithm~\ref{alg_dfs} and~\ref{alg_dfsvisit}.

\begin{algorithm} \label{alg_dfs}
	\textbf{Given:} System digraph $\mc{G}$
	
	\KwResult{Node attributes $\mc{V}_i.u$, $ \mc{V}_i.\pi$, $\mc{V}_i.s$, and $ \mc{V}_i.e$ for $i=\{1,...,n\}$}
	$t = 0 $\;
	\For{$\mc{V}_i \in \mc{V}$}{
		$\mc{V}_i.\pi = \emptyset$ \;
		$\mc{V}_i.u = \mbox{false}$\;
		$\mc{V}_i.s = 0$ \;
		$\mc{V}_i.e = 0$\;		
		}
	\For{$\mc{V}_i \in \mc{V}: \mc{V}_i.u == \mbox{false} $}{
    $\mbox{DFSvisit}(\mc{G},\mc{V}_i)$}
	\textbf{Return} $\mc{V}_i.u$, $ \mc{V}_i.\pi$, $\mc{V}_i.s$, and $ \mc{V}_i.e$\;
	\caption{This algorithm, called as \texttt{DFS}, returns the node attributes associated with every node in $\mc{G}$ calling the algorithm \texttt{DFSvisit} (Algorithm~\ref{alg_dfsvisit}).}
\end{algorithm}
\begin{algorithm} \label{alg_dfsvisit}
	\textbf{Given:} System digraph $\mc{G}$, node $\mc{V}_i$
	
	\KwResult{$\mbox{DFSvisit}(\mc{G},\mc{V}_i)$}
			 $\mc{V}_i.u == \mbox{true}$\;
			 $t = t+1$ \;
			 $\mc{V}_i.s = t$ \;
			 \For{$\mc{V}_j \in \mc{N}'(\mc{V}_i)$}{
			 	\If{$\mc{V}_j.\pi==\mbox{false}$}{
			 		$\mc{V}_j.\pi = \mc{V}_i$\;
			 		$\mbox{DFSvisit}(\mc{G},\mc{V}_j)$
			 }}
			 $t = t+1$ \;
			 $\mc{V}_i.e = t$ \;			
			 \textbf{Return} $\mbox{DFSvisit}(\mc{G},\mc{V}_i)$\;
	\caption{This algorithm has been called in Algorithm~\ref{alg_dfs} as \texttt{DFSvisit}. Note that  $\mc{N}'(\mc{V}_i)=\{\mc{V}_j|(\mc{V}_i,\mc{V}_j) \in \mc{E}\}$.}
\end{algorithm}
These algorithms are applied to find the SCCs and particularly the child SCCs in system digraph $\mc{G}$ as summarized in the Algorithm~\ref{alg_scc}.

\begin{algorithm} \label{alg_scc}
	\textbf{Given:} System digraph $\mc{G}$
	
	\KwResult{SCCs $\mc{S}_i$ and child SCCs $\mc{S}^c_i$}
	$\mbox{DFS}(\mc{G})$ \;
	Find $\mc{V}_i.e$ for every $\mc{V}_i \in \mc{V}$ \;
	Find $\mc{G}^T$ \;
	$\mbox{DFS}(\mc{G}^T)$ with nodes $\mc{V}_i$ in decreasing order of $\mc{V}_i.e$ \;
	$\mc{S}_i$ includes all nodes in DFS forest over $\mc{G}^T$ \;
	Define $\mc{S}_i^c$ as $\mc{S}_i$ including no $\mc{V}_j$ with $\mc{V}_j.\pi$ not in $\mc{S}_i$ \;
	\textbf{Return} SCCs $\mc{S}_i$ and child SCCs $\mc{S}^c_i$\;
		\caption{This algorithm finds the SCCs and child SCCs in the system digraph $\mc{G}$ applying the \texttt{DFS} algorithm. Note that the digraph  $\mc{G}^T$ is the transpose of the system digraph $\mc{G}$ obtained by reversing the link directions in $\mc{G}$.}
\end{algorithm}

%NOte that, the partial order $\prec$ can be  defined based on the variable $\mc{V}_i.\pi$. If for SCC $\mc{S}_i$

{
\subsection{Complexity of the algorithms} \label{sec_complexity}
	The procedure of finding $\underline{\mc{M}}$-augmenting paths $\mc{P}_{\underline{\mc{M}}}$ in Algorithm~\ref{alg_dilation} is executed in time-complexity $\mc{O}(|\mc{E}|+|\mc{V}|)$. The first loop  runs $|\delta \underline{\mc{M}}|$ times which, in  worst-case scenario, 
	is of $\mc{O}(|\mc{M}|^{0.5})$. Therefore, the first loop to find the maximum matching $\mc{M}$ is   $\mc{O}((|\mc{E}|+|\mc{V}|)|\mc{M}|^{0.5})$ complexity. The complexity of the second loop of the algorithm is  similarly $\mc{O}((|\mc{E}|+|\mc{V}|)|\mc{M}|^{0.5})$. Therefore, with $|\mc{V}|=n$, the number of links in the graph $|\mc{E}|$ is at most $n(n-1)/2$, and the size of maximum matching $|\mc{M}|$ is at most $n$. This implies that  Algorithm~\ref{alg_dilation} is of time-complexity $\mc{O}(n^{2.5})$. 
}

{
The running time of the first loop in Algorithm~\ref{alg_dfs} is $\mc{O}(|\mc{V}|)$. In the next loop, Algorithm~\ref{alg_dfsvisit} is called once for each node $\mc{V}_i \in \mc{V}$. The loop in Algorithm~\ref{alg_dfsvisit} for every $\mc{V}_i \in \mc{V}$ executes $\mc{N}'(\mc{V}_i)$ times. Note that, $\sum_{\mc{V}_i \in \mc{V}} |\mc{N}'(\mc{V}_i)|$ is $\mc{O}(|\mc{E}|)$.
Therefore, the total running time of Algorithm~\ref{alg_dfsvisit} is $\mc{O}(|\mc{E}|)$. Using aggregate analysis\cite{algorithm}, the overall running time of Algorithm~\ref{alg_dfs} is $\mc{O}(|\mc{V}|+|\mc{E}|)$. Having $|\mc{V}|=n$, the complexity of Algorithm~\ref{alg_dfs} is $\mc{O}(n^2)$. Finally, Algorithm~\ref{alg_scc} runs Algorithm~\ref{alg_dfs} twice and, therefore, its time complexity is similarly $\mc{O}(n^2)$. 
}

\section{Minimal Driver Nodes in System Digraph} \label{sec_min}
Using the graph-theoretic notions in Sections~\ref{sec_prob} and \ref{sec_graph}, the main results on minimizing $|\mc{B}|_0$ for structural controllability are derived in this section.  $|\mc{B}|_0$ can be defined as the minimal number of driver nodes in the system digraph $\mc{G}$ that are injected by an input. We denote this number by $N_{\min}$ in the rest of the paper.
%\begin{lem} \label{lem_cycle}
%    Injecting an input to (at least) one node in every dilation $\mc{D}_i$ satisfies the condition (ii) in Theorem~\ref{thm_scont}.
%\end{lem}	
In the light of Theorem~\ref{thm_scont} and using the notions of dilations and child SCCs the conditions for structural controllability are redefined in the following theorem.
\begin{theorem} \label{thm_scont2}
    The necessary and sufficient conditions for structural controllability of the system digraph $\mc{G}$ associated with the dynamical system~\eqref{eq_nonlin_sys}-\eqref{eq_lin_sys} are as follows:
	\begin{enumerate} [(i)]
		\item at least one state node in every child SCC, $\mc{S}^c_i,~i=\{1,...,p\}$, must be driven by an input. This node is referred to as Type-I driver node.
		\item at least one state node in every dilation, $\mc{D}_i,~ i=\{1,...,l\}$, must be driven by an input. This node is referred to as Type-II driver node.
	\end{enumerate}
\end{theorem}
{
\begin{proof} The proof of condition (ii)	for SC system digraphs is given in \cite{Liu_nature}. 
	%The same line of proof can be easily extended to non-SC directed graphs. 
	Assuming that condition (ii) holds, condition (i) is proved for general non-SC digraphs in the following.\\
\textit{Necessity:} The necessity is proved by contradiction. Assume no node in child SCC $\mc{S}^c_i$ is driven by an input. According to condition (i) in Theorem~\ref{thm_scont}, for structural controllability every node in $\mc{S}^c_i$ must be the end node of a directed path initiated by an input. From the definition of child SCC, there is no SCC $\mc{S}_j$ such that $\mc{S}_j \prec \mc{S}^c_i$. This implies that there is no $\mc{S}_j$ with direct link or directed path from the nodes in $\mc{S}_j$ to nodes in $\mc{S}^c_i$. This along with the contradiction  assumption of no driver node in $\mc{S}^c_i$ implies that the input-connectivity condition in Theorem~\ref{thm_scont} is not satisfied and therefore the nodes in $\mc{S}^c_i$ are not controllable.
\\
\textit{Sufficiency:} Assume that there is one driver node in every child SCC $\mc{S}^c_i$ injected by an input. Based on the definition of SCC, there is a directed path from this driver node to every other node in $\mc{S}^c_i$. Further, for every parent SCC $\mc{S}_j$ there is a child SCC $\mc{S}^c_i$ such that $\mc{S}^c_i \prec \mc{S}_j$. Therefore, the input-connectivity of the nodes in $\mc{S}^c_i$ implies the input-connectivity of the nodes $\mc{S}_j$. This holds for every parent SCC $\mc{S}_j$. Assuming that the condition (ii) holds, both conditions in Theorem~\ref{thm_scont} are satisfied and the structural controllability follows.
\end{proof}
}
{Note the difference between Type-I and Type-II driver nodes in Theorem~\ref{thm_scont2}. Injecting input to Type-I driver nodes recovers the input-connectivity condition in Theorem~\ref{thm_scont}, while input to Type-II driver nodes recovers  the rank-condition. The literature, e.g. \cite{Liu_nature}, only considers Type-II driver nodes, assuming that the  network is SC and the input-connectivity in Theorem~\ref{thm_scont} holds. In this work, Type-I driver nodes are introduced for controllability analysis of general non-SC networks.}
\begin{lem} \label{lem_disjoint}
    Every two SCCs are disjoint, but two dilations may share nodes.
\end{lem}
{
\begin{proof}The SCCs being disjoint simply follows the definition. Two SCCs $\mc{S}_i$ and $\mc{S}_j$  sharing a node $\mc{V}_k$ implies that there is a directed path from all nodes in $\mc{S}_i$ to $\mc{V}_k$ and from all nodes in $\mc{S}_j$ to $\mc{V}_k$; therefore, there are directed paths between the nodes in $\mc{S}_i$ and $\mc{S}_j$, implying that these two components making a larger SCC. Note that this contradicts the definition of SCCs as the  \textit{largest} set of nodes connected via directed paths.
Further, dilations that share nodes are prevalent, e.g., in \textit{star} digraphs. 
\end{proof}
}
We refer interested readers to the previous work by the author \cite{icassp13} for rank-deficient graph examples.
%\footnote{The work \cite{icassp13} discusses the dual problem of observability. The graph examples can be easily redefined for structural controllability and dilations by reversing the links in the system digraphs.}. 
Also, it should be noted that the dilation sets and child SCCs may share nodes. Examples are given in Section~\ref{sec_example}.

\begin{lem} \label{lem_unmatched}
	For every unmatched node in $\delta \mc{M}$ there exists one dilation. In other words, $	|\delta \mc{M}|=|\mc{D}|$.
\end{lem}
{
\begin{proof}
This lemma follows the definition of dilation sets. As discussed in Section~\ref{sec_dil}, the procedure of finding a dilation set $\mc{D}_i$ starts with an unmatched node in $\delta \mc{M}$ and then finds all the reachable nodes in $\mc{V}^-$ by $\mc{M}$-alternating paths $\mc{Q}_\mc{M}$. Therefore, there is one dilation set for every unmatched node. 
\end{proof}
}
\begin{cor} \label{cor_unmatched}
	Every choice of maximum matching  $\mc{M}$ accompanies with  one unmatched node in every dilation set $\mc{D}_i$.
\end{cor}
{
\begin{proof}
The proof follows from Section~\ref{sec_dil} and the same line of justifications as in the proof of  Lemma~\ref{lem_unmatched}.
\end{proof}
}
Note that the maximum matching is not unique in general and every choice of maximum matching $\mc{M}$ results a different set of unmatched nodes $\delta \mc{M}$, while every unmatched node belongs to a dilation set $\mc{D}_i$.  In fact, the dilation sets $\mc{D}_i, i=\{1,...,l\}$ include all possible sets of unmatched nodes for different choices of maximum matching. Note that injecting input to nodes in dilation sets improves the rank condition in Theorem~\ref{thm_scont}. This is tightly related with the concept of structural-rank (or S-rank) of the system adjacency matrix (or the Jacobian matrix).
\begin{defn} \cite{murota}
Define the S-rank of $J$ as the maximum possible rank of the matrix $J$ by changing its nonzero entries. In other words, the S-rank of $J$ (or structured matrix $\mc{J}$) is the number of distinct nonzero entries of $J$ that share no rows and no columns.
\end{defn}

\begin{lem} \label{lem_srank}
	\cite{doostmohammadian2017observational,murota} Let matrix $B_{\mc{D}_i}$ denotes the input matrix associated with all the nodes in $\mc{D}_i$ as driver nodes. Then,
	\begin{equation}
	\mbox{S-rank}(J | B_{\mc{D}_i}) = \mbox{S-rank}(J)+1
	\end{equation}	
\end{lem}

The lemma implies that injecting input to any node in a dilation set $\mc{D}_i$ improves the rank condition in Theorem~\ref{thm_scont} by one; even if more than one driver nodes in $\mc{D}_i$ are injected by  input the S-rank recovery is one. This implies that all the state nodes in the same dilation set are \textit{equivalent} in terms of controllability. For more information on the equivalency relation for dual concept of observability  refer to the previous work by the author \cite{doostmohammadian2017observational}.
%It should be noted that injecting input to one driver node in a dilation recovers the S-rank by only one, even if the driver node is shared between two or more dilation sets. %(according to Lemma~\ref{lem_disjoint}).

\begin{cor} \label{cor_dilation_share}
	Let $\mc{D}_i \cap \mc{D}_j$ denote the shared nodes between two dilation sets $\mc{D}_i$ and $\mc{D}_j$, and matrix $B_{\mc{D}_i \cap \mc{D}_j}$ denotes the input matrix associated with these shared nodes as driver nodes. Then,
	\begin{equation}
	\mbox{S-rank}(J | B_{\mc{D}_i \cap \mc{D}_j}) = \mbox{S-rank}(J)+\min\{|\mc{D}_i \cap \mc{D}_j|,2\}.
	\end{equation}
\end{cor}
{
\begin{proof}
	The proof follows from Lemma~\ref{lem_unmatched} and~\ref{lem_srank} and the equivalency relation defined in \cite{doostmohammadian2017observational}. Note that injecting input to one driver node in a dilation, say $\mc{D}_i$ or $\mc{D}_j$, recovers the S-rank by only one, even if the driver node is shared between two dilation sets. If there are more than two shared nodes in $\mc{D}_i \cap \mc{D}_j$, by driving these shared nodes the S-rank recovery is only two due to equivalency relation.  
\end{proof}
}
This corollary can be easily generalized for more than two shared dilation sets.
In the light of Theorem~\ref{thm_scont2} and Lemma~\ref{lem_disjoint} and~\ref{lem_srank} and relevant corollaries the main result of this section is described in the following theorem.

\begin{theorem} \label{thm_min}
	The minimal number of driver nodes for structural controllability of system digraph $\mc{G}$ associated with the system models~\eqref{eq_nonlin_sys}-\eqref{eq_lin_sys} is equal to $N_{\min}=|\mc{D}|+|\mc{S}^c|-\min(|\mc{D} \cap \mc{S}^c|)$, where $\min(|\mc{D} \cap \mc{S}^c|)$ represents the minimum number of child SCCs and dilation sets that share nodes.
\end{theorem}
{
\begin{proof}
Following the results of Theorem~\ref{thm_scont2}, injecting input to one node in every dilation set and one node in every child SCC recovers the structural controllability of system digraph $\mc{G}$. From Lemma~\ref{lem_disjoint} the child SCCs do not share nodes, and further,  driving the share nodes among two or more dilation sets recovers the S-rank of the Jacobian matrix $J$ by at most the number of those dilation sets. In other words, from Lemma~\ref{lem_srank}, choosing one driver node from every dilation sets, even if the node is shared between two dilation sets, recovers the S-rank of $J$ by only one. The key point is that the dilation sets and child SCCs may share nodes (denoted by $\mc{D} \cap \mc{S}^c$); injecting inputs to these nodes recovers both conditions on input-connectivity and S-rank recovery in Theorem~\ref{thm_scont}. Therefore, choosing the nodes in $\mc{D} \cap \mc{S}^c$ as driver nodes and choosing one node from every remaining dilation sets and child SCCs renders minimal number of nodes for structural controllability.  Choosing less driver nodes than $N_{min}$ implies that there is at least one dilation set or child SCC not affected by an input which, according to Theorem~\ref{thm_scont2}, makes the digraph $\mc{G}$ structurally  uncontrollable. 
\end{proof}
}
Note that, as a result of this theorem, the minimal number of driver nodes is less than or equal to the number of child SCCs and dilation sets, i.e. $|\mc{D}| \leq N_{min} \leq |\mc{D}|+|\mc{S}^c|$. This is an improvement over the results given by \cite{Liu_nature}.

\subsection{Systems with inaccessible state nodes}
In real applications some of the state nodes might be inaccessible to be affected by input.
% This might be due to harsh environmental condition, high energy consumption,   vulnerability to cyber-attack, and even unaffordable cost to drive the state by an input. 
In this subsection, such set of nodes, denoted by $\mc{F}$, is considered which cannot be selected as  driver nodes. For such nodes the actuator (or sensor in case of observability) may fail to affect the state node \cite{doostmohammadian2017distributed,jafari2011leader}, and therefore a node in $\mc{F}$ must be avoided as a driver node. {Similar concept is discussed in \cite{shang2018resilient}, where it is considered that some of the nodes are subject to failure/attack and become dysfunctional or the nodes are non-cooperative and share wrong information over the consensus network (also known as \textit{Byzantine node}). In \cite{shang2018resilient}, the concept of resilient consensus is proposed as a counter-measure to deal with such nodes. Here, the idea of control equivalency is considered as a counter-measure to deal with inaccessible nodes. }

\begin{lem} \label{lem_inaccessible}
	Consider the digraph $\mc{G}$ associated with the system \eqref{eq_nonlin_sys}-\eqref{eq_lin_sys}. In case of having inaccessible state nodes $\mc{F}$, we have  $N_{\min}=|\mc{D}|+|\mc{S}^c|-\min(|\mc{D} \cap \mc{S}^c|-|\mc{D} \cap \mc{S}^c \cap \mc{F}|)$, where $\min(|\mc{D} \cap \mc{S}^c|-|\mc{D} \cap \mc{S}^c \cap \mc{F}|)$ denotes the minimal  number of child SCCs and dilation sets that share \textit{accessible} nodes.
\end{lem}
{
\begin{proof}
Note that in the above lemma it is assumed that at least one node in every dilation set and child SCC is accessible. Otherwise, the structural controllability cannot be achieved and the problem has no solution. As mentioned in Theorem~\ref{thm_min}, driving the  state node $\mc{V}_k$ shared between a dilation set $\mc{D}_i$ and a child SCC $\mc{S}^c_j$ recovers both conditions for structural controllability and minimizes the overall number of driver nodes. If $\mc{V}_k \in \mc{F}$ we consider two cases; case (i), assume there is no other shared node between $\mc{D}_i$ and $\mc{S}^c_j$; then, two new driver nodes including  one equivalent driver node in $\mc{D}_i$ and one equivalent driver node in $\mc{S}^c_j$ are needed to replace $\mc{V}_k$ to recover structural controllability. This implies that $N_{\min}$ is increased by one for the  inaccessible shared node in $|\mc{D} \cap \mc{S}^c \cap \mc{F}|$. For case (ii), assume there is another \textit{accessible}  node $\mc{V}_d$ shared between $\mc{D}_i$ and $\mc{S}^c_j$; then, selecting $\mc{V}_d$ as driver node recovers structural controllability.
Therefore, from these two cases, $N_{\min}$ is determined  by  the minimal number of  child SCCs and dilation sets sharing \textit{accessible}  nodes, denoted by $\min(|\mc{D} \cap \mc{S}^c|-|\mc{D} \cap \mc{S}^c \cap \mc{F}|)$.
\end{proof}
}
\subsection{Polynomial-order complexity}
For large-scale controllability applications it is preferred to apply polynomial-order algorithms. { As mentioned in Section~\ref{sec_introduction}, the related literature on minimal controllability (or dual problem of minimal observability) claim this problem to be NP-hard \cite{tzoumas2015minimal, summers2014optimal, pequito2015complexity,olshevsky2014minimal, jiang2003optimal,pequito_gsip}. Therefore, approximate solution or greedy approach is proposed to solve this problem which is not practical for large-scale application. However, as discussed in Section~\ref{sec_complexity}, the algorithms in this paper are of polynomial-order complexity. The Algorithm~\ref{alg_dilation} to find the dilation sets  is of $\mc{O}(n^{2.5})$-complexity and the overall complexity of the Algorithm~\ref{alg_scc} to find the child SCCs is  $\mc{O}(n^{2})$, with $n$ as the system size (or number of the state nodes). This implies overall complexity of order $\mc{O}(n^{2.5})$, which is significant as the computational complexity of numerical Gramian-based analysis is at least $\mc{O}(n^{3})$. 
}
%Low computational complexity of the structural approach makes the adopted algorithms preferable for analysis of large-scale system applications.

\section{Illustrative Examples}\label{sec_example}
%In this section examples are provided to illustrate the graph-theoretic notions and  results of Sections~\ref{sec_graph} and \ref{sec_min}.

\begin{figure} 
	\centering
	\includegraphics[width=2.8in]{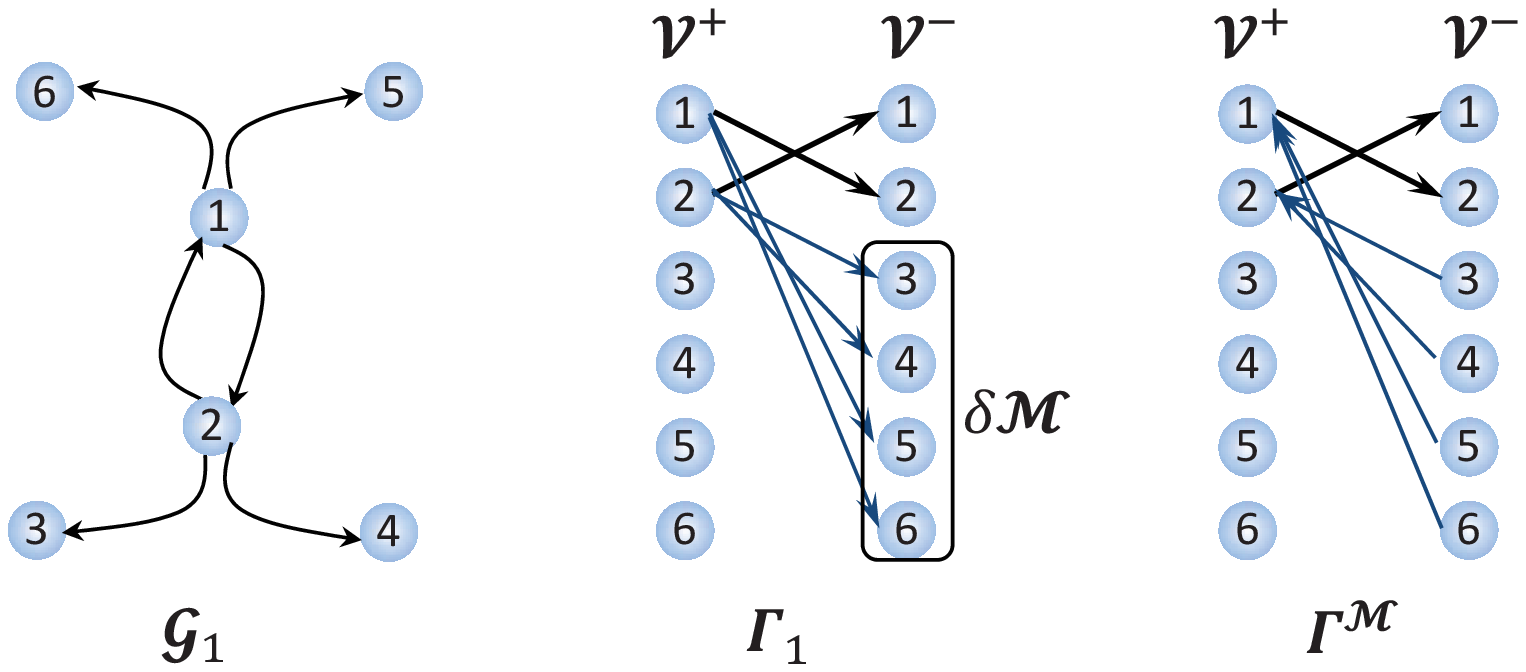}
	\caption{An example system digraph $\mc{G}_1$ is shown along with its bipartite representation $\Gamma_1$ and auxiliary graph  representation $\Gamma^\mc{M}$. The links associated with maximum matching $\mc{M}$ are shown by black and the unmatched nodes are shown as $\delta \mc{M}$.}
	\label{fig_graph1}
\end{figure}

\textit{Example 1:} Consider the system digraph $\mc{G}_1$ represented in Fig.~\ref{fig_graph1}.
This graph represents a dynamical system in the form \eqref{eq_nonlin_sys} (or the linearized model \eqref{eq_lin_sys}). For example, a link $(\mc{V}_1,\mc{V}_2)$ in $\mc{G}_1$ implies that  $\dot{x}_1$ is a function of state $x_2$, i.e. $J_{12}= \frac{\partial f_1}{\partial x_2}$ is nonzero for all operating points. Similar statement holds for all the links in the system digraph $\mc{G}_1$. Following the definitions in Section~\ref{sec_scc}, the node sets $\mc{S}_1=\{\mc{V}_1,\mc{V}_2\}$ make the only largest component in which all nodes are inter-connected via a path, and therefore $\mc{S}=\{\{\mc{V}_1,\mc{V}_2\}\}$. Further, we have $\mc{V}_1.\pi=\mc{V}_2$ and $\mc{V}_2.\pi=\mc{V}_1$ implying that this SCC has no incoming links from the nodes not belonging to itself, and therefor, it is the (only) child SCC $\mc{S}^c_1=\{\mc{V}_1,\mc{V}_2\}$.

\begin{figure} 
	\centering
	\includegraphics[width=3.3in]{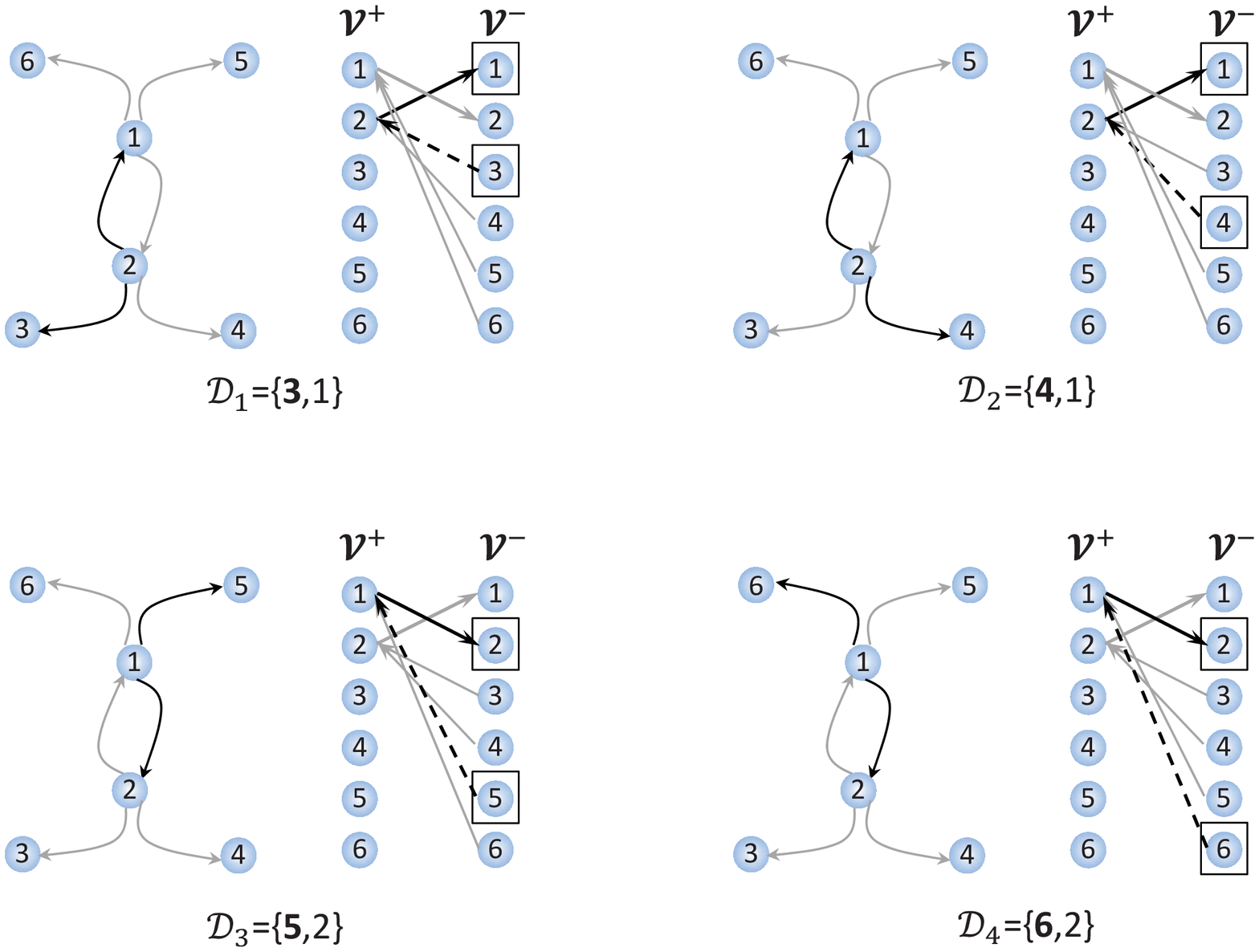}
	\caption{The dilations and dilation sets of the graph $\mc{G}_1$ (Fig.~\ref{fig_graph1}) are shown in this figure. The black links represent the dilation in the system digraph $\mc{G}_1$ and the $\mc{M}$-alternating paths in the auxiliary graph $\Gamma^\mc{M}$.} 
	\label{fig_dilation1}
\end{figure}

Next, the dilation sets are found using the graph-theoretic notions in Section~\ref{sec_dil}. The bipartite representation of graph $\mc{G}_1$ is shown in Fig.~\ref{fig_graph1} as $\Gamma_1=(\mc{V}^+,\mc{V}^-,\mc{E}_\Gamma)$. A possible  matching $\mc{M}=\{(\mc{V}_1,\mc{V}_2),(\mc{V}_2,\mc{V}_1)\}$ is shown by black links. The matched nodes incident to links in $\mc{M}$ are $\partial \mc{M}^-=\{\mc{V}_1,\mc{V}_2\}$ and the unmatched nodes are $\delta\mc{M}=\{\mc{V}_3,\mc{V}_4,\mc{V}_5,\mc{V}_6\}$. The Auxiliary graph associated with matching $\mc{M}$ is shown in Fig.\ref{fig_graph1} as $\Gamma^\mc{M}$ obtained by reversing the links not belonging to $\mc{M}$. The dilation sets are obtained by finding the nodes reachable from unmatched nodes by $\mc{M}$-alternating paths, as shown in Fig.\ref{fig_dilation1}.
The dilation sets are $\mc{D}=\{\{\mc{V}_3,\mc{V}_1\},\{\mc{V}_4, \mc{V}_1\},\{\mc{V}_5,\mc{V}_2\},\{\mc{V}_6,\mc{V}_2\}\}$. As shown in the figure, for each dilation $\mc{D}_i$ from the definition we have $|\mc{N}(\mc{D}_i)|<|\mc{D}_i|$. For example, $\mc{N}(\mc{D}_1) =\{\mc{V}_2\}$, $\mc{N}(\mc{D}_2) =\{\mc{V}_2\}$, $\mc{N}(\mc{D}_3) =\{\mc{V}_1\}$, and $\mc{N}(\mc{D}_4) =\{\mc{V}_1\}$.
To find the minimal number of driver nodes for structural controllability, following Theorem~\ref{thm_min}, $\mc{D} \cap \mc{S}^c=\{\{\mc{V}_1,\mc{V}_2\}\}$ which belongs to child SCC $\mc{S}^c_1$. Therefore,  minimal number of driver nodes is $N_{\min}=|\mc{D}|+|\mc{S}^c|-\min(|\mc{D} \cap \mc{S}^c|)=4+1-1=4$.
%Therefore, possible minimal sets of  driver nodes are, for example, $\{\mc{V}_1,\mc{V}_4,\mc{V}_5,\mc{V}_6\}$ or $\{\mc{V}_2,\mc{V}_3,\mc{V}_4,\mc{V}_6\}$.
One possible minimal set of  driver nodes is, for example, $\{\mc{V}_1,\mc{V}_4,\mc{V}_5,\mc{V}_6\}$ and the (structured) input matrix is:
\begin{eqnarray} \label{eq_input}
\mc{B} = \left(\begin{array}{cccc}
\times&0&0&0\\
0&0&0&0\\
0&0&0&0\\
0&\times&0&0\\
0&0&\times&0\\
0&0&0&\times
\end{array}
\right)
\end{eqnarray}
where $\times$ represents a nonzero entry and $0$ represents a fixed zero. The structural results match the Gramian-based numerical analysis. {For random numerical realizations of the Jacobian matrix $J$ associated with system digraph $\mc{G}_1$ and the input matrix \eqref{eq_input} the Gramian $(B|JB|J^2B|...|J^5B)$ is checked to be full row rank, verifying the structural results in this paper.}

{
\textit{Example 2:} An example system digraph $\mc{G}_2$ with $16$ state nodes similar to the social network example in \cite{pequito_gsip} is shown in Fig.\ref{fig_graph2}. 
\begin{figure} 
	\centering
	\includegraphics[width=2.8in]{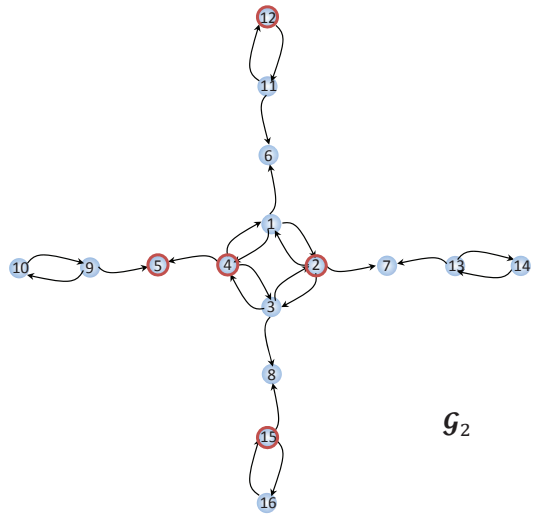}
	\caption{{An example system digraph $\mc{G}_2$ similar to the social network example in \cite{pequito_gsip}  is shown in this figure. The nodes bordered with red circle represent the inaccessible nodes that cannot be affected by a control input.}}
	\label{fig_graph2}
\end{figure}
Assume there are some inaccessible state nodes $\mc{F}=\{\mc{V}_2, \mc{V}_4, \mc{V}_5, \mc{V}_{12}, \mc{V}_{15}\}$ in the system which cannot be affected by input. Using Algorithm~\ref{alg_dilation} and Algorithm~\ref{alg_scc} the dilation sets and child SCCs are found in the system digraph as follows:  $\mc{D}=\{\mc{D}_1,\mc{D}_2,\mc{D}_3,\mc{D}_4 \}$ where $\mc{D}_1= \{\mc{V}_1,\mc{V}_3,\mc{V}_5,\mc{V}_{10}\}$, $\mc{D}_2 = \{\mc{V}_2,\mc{V}_4,\mc{V}_6,\mc{V}_{12}\}$, $\mc{D}_3=\{\mc{V}_1,\mc{V}_3,\mc{V}_7,\mc{V}_{14}\}$, $\mc{D}_4=\{\mc{V}_2,\mc{V}_4,\mc{V}_8,\mc{V}_{16}\}$, and $\mc{S}^c=\{\mc{S}^c_1,\mc{S}^c_2,\mc{S}^c_3,\mc{S}^c_4,\mc{S}^c_5\}$ where $\mc{S}^c_1=\{\{\mc{V}_1,\mc{V}_2,\mc{V}_3,\mc{V}_4\}$, $\mc{S}^c_2=\{\mc{V}_9,\mc{V}_{10}\}$, $\mc{S}^c_3=\{\mc{V}_{11},\mc{V}_{12}\}$, $\mc{S}^c_4=\{\mc{V}_{13},\mc{V}_{14}\}$, $\mc{S}^c_5 = \{\mc{V}_{15},\mc{V}_{16}\}$. To find the minimal driver nodes, $\mc{D} \cap \mc{S}^c=\{ \{\mc{V}_1,\mc{V}_3,\mc{V}_{10}\}, \{\mc{V}_2,\mc{V}_4,\mc{V}_{12}\}, \{\mc{V}_1,\mc{V}_3,\mc{V}_{14}\}, \{\mc{V}_2,\mc{V}_4,\mc{V}_{16}\}\}$ and, therefore, following Theorem~\ref{thm_min} $N_{\min}=|\mc{D}|+|\mc{S}^c|-\min(|\mc{D} \cap \mc{S}^c|)=5+4-4=5$.  One possible set of driver nodes is $\{\mc{V}_1, \mc{V}_{10},\mc{V}_{12},\mc{V}_{14},\mc{V}_{16}\} $. This is by assumption that all nodes are accessible. However, considering the inaccessible nodes as in Fig.\ref{fig_graph2}, $\mc{D} \cap \mc{S}^c \cap \mc{F}=\{\{\mc{V}_2,\mc{V}_4,\mc{V}_{12}\}\}$ and, therefore, following Lemma~\ref{lem_inaccessible}
$N_{\min}=|\mc{D}|+|\mc{S}^c|-\min(|\mc{D} \cap \mc{S}^c|-|\mc{D} \cap \mc{S}^c \cap \mc{F}|)=5+4-3=6$.  Equivalent accessible nodes in the same child SCC $\mc{S}^c_3$ and dilation set $\mc{D}_2$ replace the node $\{\mc{V}_{12}\}$ for controllability recovery. One possible set of driver nodes is $\{\mc{V}_1,\mc{V}_6, \mc{V}_{10},\mc{V}_{11},\mc{V}_{14},\mc{V}_{16}\} $. Having the controllability Gramian to be full row-rank verifies the structural results.
%with  the structured input matrix as:
%\begin{eqnarray} \label{eq_input1} \tiny
%\mc{B} = \left(\begin{array}{cccccccccccccccc}
%\times&0&0&0&0&0&0&0&0&0&0&0&0&0&0&0\\
%0&0&0&0&0&\times&0&0&0&0&0&0&0&0&0&0\\
%0&0&0&0&0&0&0&0&0&\times&0&0&0&0&0&0\\
%0&0&0&0&0&0&0&0&0&0&\times&0&0&0&0&0\\
%0&0&0&0&0&0&0&0&0&0&0&0&0&\times&0&0\\
%0&0&0&0&0&0&0&0&0&0&0&0&0&0&0&\times
%\end{array}
%\right)^\top
%\end{eqnarray}
%Note the transpose sign in the above equation. 
%It is worth noting that if all nodes in a dilation set or a child SCC are inaccessible, for example $ \mc{F}=\{\mc{V}_2,\mc{V}_4,\mc{V}_6,\mc{V}_{12}\}$, the structural controllability has no solution.
}
\section{Concluding Remarks}\label{sec_con}

Results of this paper find applications in network medicine. For example, to find the driver nodes in a gene regulatory network \cite{muller2011few} or to minimally regulate the cell functions via  cellular differentiation process \cite{rajapakse2012can}. Further, the dual problem of minimal observer nodes to monitor power grid \cite{camsap11,heussen2011unified,haynes2002domination,kneis2006parameterized} is another direction of future research. Another open problem is in the minimal controllability of composite networks made by graph product of factor networks \cite{chapman2014controllability}. 
In the same line of research, an open problem is the minimal link addition and topological change for controllability recovery to reduce the driver nodes in complex networks. Following \cite{alcaraz2013structural}, investigating the effect of node removal on minimal driver nodes  in ER, SW, and SF networks is another future research direction. Thus, one can compare the number of driver nodes in different types of these networks and also investigate the effect of changing their parameters on number of driver nodes.

\bibliographystyle{IEEEbib}
\bibliography{bibliography}

%\begin{IEEEbiography}[{\includegraphics[width=1.1in]{bio_me.eps}}]{Mohammadreza~Doostmohammadian}
%	received his B.Sc. and M.Sc. in Mechanical Engineering from Sharif University of Technology, Tehran, Iran, respectively in 2007 and 2010, where he worked on different applications of control systems and robotics. He received his PhD in Electrical Engineering from Tufts University, Medford, USA in 2015. During his PhD at Signal Processing and Robotic Network (SPARTN) lab he worked on control and signal processing over networks with particular application in social networks. From 2015 to 2017 he was a post-doctoral researcher at ICT Innovation Center for Advanced Information and Communication Technology (AICT), School of Computer Engineering, Sharif University of Technology, where he focused his research on control and estimation over networks and network epidemic. He was a researcher at Iran Telecommunication Research Center (ITRC), Tehran, Iran in 2017, working on estimation over IoT. Currently, he is an Assistant Professor of Mechatronics at Semnan University, Semnan, Iran. His general research interests include control, estimation, and complex networks. He was the chair of robotics and control session at ISME-2018 conference. He is a reviewer for IFAC and IEEE journals and conferences.
%\end{IEEEbiography}

\end{document}